\documentclass[aps,12pt,preprintnumbers,nofootinbib]{revtex4}

\usepackage{graphicx}
\usepackage{amsmath,amssymb,epsfig,bm,comment}
\usepackage{fancyhdr}
\usepackage[dvips]{color}
\usepackage{mathrsfs}

\definecolor{gray}{rgb}{0.4,0.4,0.4}

\def\({\left(}
\def\){\right)}
\def\[{\left[}
\def\]{\right]}
\def\<{\langle}
\def\>{\rangle}

\newcommand{\be}{\begin{equation}}
\newcommand{\ee}{\end{equation}}
\newcommand{\bea}{\begin{eqnarray}}
\newcommand{\eea}{\end{eqnarray}}
\newcommand{\bwt}{\begin{widetext}}
\newcommand{\ewt}{\end{widetext}}

\newcommand{\bi}{\begin{itemize}}
\newcommand{\ei}{\end{itemize}}
\newcommand{\ben}{\begin{enumerate}}
\newcommand{\een}{\end{enumerate}}
\newcommand{\bca}{\begin{cases}}
\newcommand{\eca}{\end{cases}}
\newcommand{\bln}{\begin{align}}
\newcommand{\eln}{\end{align}}
\newcommand{\bst}{\begin{split}}
\newcommand{\est}{\end{split}}

\newcommand{\tf}{\widetilde{F}}
\newcommand{\ce}{\mathcal E}
\newcommand{\cm}{\mathcal M}
\newcommand{\ck}{\mathcal K}
\newcommand{\cl}{\mathcal L}
\newcommand{\ch}{\mathcal H}
\newcommand{\cg}{\mathcal G}
\newcommand{\httc}{{h^t_t}^{(0)}}
\newcommand{\hxxc}{{h^x_x}^{(0)}}
\newcommand{\hzzc}{{h^z_z}^{(0)}}
\newcommand{\htzc}{{h^z_t}^{(0)}}
\newcommand{\btc}{{B_t}^{(0)}}
\newcommand{\bzc}{{B_z}^{(0)}}

\begin{document}
\preprint {KEK-TH-1543}
\title{Holographic RG  flow and  Sound Modes of sQGP }

\author{Yoshinori~Matsuo$^{\ddagger}$, Sang-Jin~Sin$^{\dagger}$ and Yang~Zhou$^{*}$}
\affiliation{
$\ddagger$ High Energy Accelerator Research Organization(KEK),
Tsukuba, Ibaraki 305-0801, Japan\\
$\dagger$ Department of Physics, Hanyang University,
Seoul 133-791, Korea\\
$*$ Center for Quantum
Spacetime, Sogang University, Seoul 121-742, Korea}
\begin{abstract}
We consider the hydrodynamics of strongly interacting quark gluon plasma in finite temperature and density
using the  holographic duality of charged black hole in anti DeSitter space. 
We calculate the transport coefficients at arbitrary energy scale by considering the holographic screen at finite radial position. 
We first calculate the flow of  sound velocity in this method and check the consistence with previous result. 
Then we  calculate diffusion constant of charge and find that Einstein relation between susceptibility, conductivity and diffusion constants  will hold 
at arbitrary slice.
\end{abstract}
\maketitle
\section{Introduction}

In usual application of AdS/CFT~\cite{Maldacena:1997re}, physical quantities are calculated at the  asymptotic  AdS boundary. However,  a single asymptotical AdS gravity system is expected to describe a theory   both at the UV regime  as well as the IR region,
 since the radial direction is identified  as the   energy scale~\cite{kraus, verlinde, Susskind} of the gauge theory. 
While the asymptotic boundary describes the UV fixed point,  the horizon, as IR boundary, should be able to characterize   low energy behavior of the theory. 
 In fact, some of the transport coefficients were calculated using membrane paradigm at the horizon.  
  Therefore it would be very interesting to find the interpolating functions connecting the 
  UV and the IR behavior.  
  Such interpolating functions can be computed by putting the holographic screen at a slice $r=r_c$. 
   The paradigm to compute the slice dependent quantities  has been recently proposed in~\cite{Joe.RG, Hong.RG} as holographic Wilsonian renormalization group.  Physical  quantities at the slice have been defined in a slight different way of Wilson fluid/gravity duality in~\cite{Andy.RG}. It has been shown that these two different methods, together with the earlier sliding membrane paradigm~\cite{Hong.Membrane} are consistent~\cite{SZ,MSZ}. For further recent progress in this direction, see~\cite{Laia:2011wf,Elander:2011vh,Heemskerk:2012mn, Marolf:2012dr, Bousso:2012sj,Iqbal:2011ae,Lee:2012xb}.

When we ask the density dependence, we need to introduce the local U(1) charge, which in turn 
gives an extra complexity in the hydrodynamic analysis due to the mode mixing:   
  in the RN-AdS black hole,  vector modes of gravitational fluctuations   mix with transverse Maxwell modes
  \cite{KI,Ge:2008ak}. The scalar modes of metric perturbation mix with longitudinal modes of Maxwell potential and the analysis is much more involved~\cite{MSTTY}. 
  In a companion paper~\cite{MSZ}, we focused on the former, i.e, the  holographic RG flow 
of tensor and vector part of metric perturbations and transverse Maxwell perturbation. We proposed a systematical method to compute the hydrodynamical poles and retarded Green function at the slice  $r=r_c$. 

In this paper, we   focus on the running of the  sound  modes. Notice that 
the running of sound modes has already been studied in~\cite{Andy.RG} and  in more recent paper~\cite{Marolf:2012dr}.   What is new in this paper is the density effect on the sound modes coming from the mixing of gravity and Maxwell part in the presence of the charge. We will obtain    analytical results for  sound velocity and  charge-diffusion constant and will see that they have nontrivial charge dependence in RG flows.   We also 
find that Einstein relation between susceptibility, conductivity and diffusion constants  will hold 
at arbitrary slice.

This paper is organized as follows. In section \ref{Smode}, we analyze the sound part perturbations for the RN-AdS$_5$ and focus on the dynamical structure of Einstein and Maxwell equations.
In section \ref{Meom}, we write down the master equations which are equivalent to equations of motion in section \ref{Smode} and solve them perturbatively in hydrodynamical limit. We impose both the in going horizon condition and Dirichlet boundary conditions at the slice surface $r=r_c$. In the following section \ref{pole} we derive the on shell action for the sound fluctuations and write down forms of Green functions and pick up the pole position. We found a charge diffusion pole and a sound pole at $r=r_c$ and we discuss the RG flow of them. We found the {\it universal} value (charge independent) located at the horizon for both momentum diffusion and charge diffusion constant while located at the UV boundary for the speed of sound. We checked some transport coefficients and Einstein relations at the slice in the last section.

\section{Sound Modes in Reissnner-Nordstr\"om-AdS$_5$}\label{Smode}

We consider the 5 dimensional Einstein-Maxwell theory. 
The action is given by 
\begin{equation}
 S = \frac{1}{2\kappa^2}\int d^5 x \sqrt{-g} 
  \left(R - 2\Lambda\right) 
  - \frac{1}{4e^2} \int d^5 x \sqrt{-g}\, F^2 , 
\end{equation}
where five dimensional Newton constant is given by $G_5=\kappa^2/ 8\pi $. 
The metric of the RN-AdS black hole is given by 
\begin{equation}
 ds^2 = \frac{r^2}{l^2}
  \left(-f(r) dt^2 + d\vec x^2 \right) + \frac{l^2 dr^2}{r^2 f(r)} , 
\end{equation}
where 
\begin{align}\label{RNmetric}
 f(r) &= 1 - \frac{l^2 m}{r^4}+ \frac{l^2 q^2}{r^6} , &
 \Lambda &= - \frac{6}{l^2} , 
\end{align}
and the background gauge potential is 
\begin{equation}\label{RNgauge}
 A_t = \mu - \frac{Q}{r^2} , 
\end{equation}
where 
\begin{equation}
 \frac{Q^2}{e^2} = \frac{3 q^2}{2 \kappa^2} . 
\end{equation}
The chemical potential $\mu$ is fixed by the regularity condition 
at the horizon as 
\begin{equation}
 \mu = \frac{Q}{r_+^2} . 
\end{equation}
It is convenient to introduce a coordinate $u = r_+^2/r^2$, 
where $r_+$ is the outer horizon. 
Then the metric is expressed as 
\begin{align}
 ds^2 &= \frac{l^2}{4b^2 u}\left(-f(u)dt^2 + d \vec x^2\right) 
  + \frac{l^2 du^2}{4 u^2 f(u)} , \\
 & f(u) = (1-u) (1+u-a u^2) , 
\end{align}
where 
\begin{align}
 a &= \frac{l^2  q^2}{r_+^6} , & 
 b &= \frac{l^2}{2r_+} . 
\end{align} Consider the gravitational fluctuation for the RN-AdS solution, the Gibbons-Hawking term on the boundary is necessary to well define a variation principle
\begin{equation}
 S_\text{GH} = \frac{1}{\kappa^2} \int d^4 x \sqrt{-\gamma}\,K 
\end{equation}
where $\gamma_{\mu\nu}$ and $K_{\mu\nu}$ are 
induced metric and extrinsic curvature on the boundary, respectively. Furthermore, 
in order to regularize the on shell action, 
we introduce the counter term~\cite{Balasubramanian:1999re}
\begin{equation}
 S_\text{ct} = \frac{1}{\kappa^2}\int d^4 x \sqrt{-\gamma} \frac{3}{l} . 
\end{equation}

From now on, we study fluctuations $h_{mn},\ B_m$ for 
the background metric $\bar g_{mn}$ (\ref{RNmetric}) and the gauge field $\bar A_m$ (\ref{RNgauge}), 
\begin{align}
 g_{mn} &= \bar g_{mn} + h_{mn} , & 
 A_m &= \bar A_m + \frac{B_m}{\mu} , 
\end{align}
and consider the linearized theory of these fluctuations. 
Take the gauge condition
$h_{um} = 0$ and $B_u = 0$ and consider the Fourier transforms of these fluctuation fields, 
\begin{align}
 h_{\mu\nu}(x) 
 &= \int \frac{d^4k}{(2\pi)^4} 
 e^{-i\omega t + i k z} h_{\mu\nu}(r)\ , \\\
 B_\mu 
 &= \int \frac{d^4k}{(2\pi)^4} 
 e^{-i\omega t + i k z} B_\mu(r)\ ,
\end{align}
where the
momenta $k$ is along the $z$-direction. 
The fluctuations of the metric are categorized into 
the tensor mode, shear mode and sound mode, and 
the gauge field is categorized into 
the vector mode and scalar mode. 
At the linearized order, 
there are mixture between the shear part of metric and 
transverse gauge field, also
the sound mode of metric and the longitudinal gauge field. 
Here, we focus on the sound mode of the metric and the vector mode 
of the gauge field. 
Then, the relevant components are 
\begin{equation}
 h^t_t,\qquad 
 h^x_x=h^y_y,\qquad 
 h^z_z,\qquad 
 h^z_t,\qquad 
 B_t,\qquad 
 B_z
\end{equation}
where we have defined $h^\mu_\nu = \bar g^{\mu\lambda} h_{\lambda\nu}$
for later convenience. 

Relevant equations for the sound modes in the Einstein equation 
are 
$(t, t)$, $(t, u)$, $(x, x)$, $(z, z)$, $(u, u)$, 
$(t, z)$ and $(u, z)$ components listed as follows: 
\begin{subequations}
 \begin{align}
  0
  &=
  {h^t_t}'' +\frac{(u^{-1}f)'}{u^{-1}f}
  \left(\frac{3}{2}{h^t_t}'+{h^x_x}'+\frac{1}{2}{h^z_z}'\right)
  -\frac{b^2k^2}{uf}h^t_t 
\notag\\&\quad
  +\frac{2b^2}{uf^2}
  \left(
  \omega^2h^x_x +\frac{1}{2}\omega^2h^z_z +\omega kh^z_t
  \right)
  +2a\frac{u}{f}h^t_t +4a\frac{u}{f}B_t', \label{Ett}
  \\
  0
  &=
  {h^x_x}'' +\frac{(u^{-2}f)'}{u^{-2}f}{h^x_x}'
  -\frac{1}{2u}\left({h^t_t}'+{h^z_z}' \right)
  +\frac{b^2}{uf^2} \left( \omega^2-k^2f \right)h^x_x
  \notag 
  \\
  &\quad
  -a\frac{u}{f}h^t_t -2a\frac{u}{f}B_t', 
  \label{Exx}
  \\
  0
  &=
  {h^z_z}'' +\frac{(u^{-\frac{3}{2}}f)'}{u^{-\frac{3}{2}}f}{h^z_z}'
  -\frac{1}{u} \left(\frac{1}{2}{h^t_t}'+{h^x_x}' \right)
  +\frac{b^2}{uf^2}
  \left( \omega^2h^z_z +2\omega kh^z_t -k^2fh^t_t -2k^2fh^x_x \right)
  \notag 
  \\
  &\quad
  -a\frac{u}{f}h^t_t -2a\frac{u}{f}B_t', 
  \label{Ezz}
  \\
  0
  &=
  {h^z_t}'' -\frac{1}{u}{h^z_t}' +\frac{2b^2\omega k}{uf}h^x_x -3auB_z', 
  \label{Etz}
  \\
  0
  &=
  {h^t_t}'' +2{h^x_x}'' +{h^z_z}'' +\frac{f'}{f}
  \left( \frac{3}{2}{h^t_t}' +{h^x_x}' +\frac{1}{2}{h^z_z}' \right)
  +2a\frac{u}{f}h^t_t +4a\frac{u}{f}B_t', 
  \label{Euu}
  \\
  0
  &=
  \omega
  \left[
  2{h^x_x}'+{h^z_z}' -\frac{f'}{f} \left(h^x_x+\frac{1}{2}h^z_z \right)
  \right]
  +k
  \left(
  {h^z_t}'-\frac{f'}{f}h^z_t
  \right), 
  \label{Etu}
  \\
  0
  &=
  k{h^t_t}' +2k{h^x_x}' -\frac{\omega}{f}{h^z_t}' +\frac{kf'}{2f}h^t_t
  +3a\frac{u}{f} \left( kB_t+\omega B_z \right)\ .
  \label{Ezu}
 \end{align}
\end{subequations}
The Maxwell equations which is relevant 
to the vector mode are $t$, $z$, and $u$ components: 
\begin{subequations}
 \begin{align}
  0
  &=
  B_t'' -\frac{b^2}{uf} \left( k^2B_t+k\omega B_z \right)
  +\frac{1}{2} \left( {h^t_t}'-2{h^x_x}' -{h^z_z}' \right), 
  \label{Mt}
  \\
  0
  &=
  B_z'' +\frac{f'}{f}B_z' +\frac{b^2}{uf^2}
  \left( \omega^2B_z+\omega kB_t \right) -\frac{1}{f} {h^z_t}',  
  \label{Mz}
  \\
  0
  &=
  \omega B_t' +kf B_z' +\frac{\omega}{2} \left( h^t_t-2h^x_x-h^z_z \right)
  -kh^z_t. 
  \label{Mu}
 \end{align}
\end{subequations}
There are 10 equations for 6 fields, and only 6 of 
these equations are independent. 
These equations are classified into 
the dynamical equations which are 
second order differential equations 
and first order constraints. 
For the Einstein equations, 
we have 4 dynamical equations and 3 constraint equations. 
Independent equations are 1 dynamical equation and 3 constraint equations. 
For the Maxwell equations, 
1 dynamical equation and 1 constraint equation are independent. 
These differential equations give 8 integration constants
since the solutions of $n$-th differential equations have 
$n$ integration constants. 
We impose the incoming boundary condition at the horizon 
for each dynamical equation. 
Then, there remain 6 integration constants 
which are determined by choosing the values of 
the 6 fields on the UV boundary. 

\section{Master equations and Boundary Conditions}\label{Meom}
\subsection{Master equations}
Consider the two independent dynamical equations. 
By using an appropriate redefinition of the fields, 
we obtain the dynamical equations which always 
satisfy the constraint equations. 
Such fields and equations are called master fields and master equations, respectively. 
The master fields are derived in \cite{KI}, 
and can be expressed in our notation as 
\begin{equation}\label{mas1}
 \Phi = \frac{1}{4u^{3/2} (4 b^2 k^2 - 3 f')}
  \left[
   (4 b^2 k^2 - 3 f') h^x_x 
   + 2 f \left({h^x_x}' + {h^z_z}'\right)
  \right] , 
\end{equation}
for metric perturbations and 
\begin{equation}\label{mas2}
 \mathcal A = 2 a 
  \left(
   - h^t_t + 3 h^x_x - 2 B_t'
  \right) , 
\end{equation}
for gauge part. 
Two master equations for the master fields (\ref{mas1}) and (\ref{mas2}) can be
diagonalized by the following new master fields: 
\begin{equation}
 \Phi_\pm = \alpha_\pm \Phi + \frac{u^{1/4}}{8} \mathcal A , 
\end{equation}
where 
\begin{align}
 \alpha_\pm &= (1+a)(1\pm\alpha) -3 a u_c , \\
 \alpha &= \sqrt{1 + \frac{4 a b^2 k^2}{(1+a)^2}}\ .
\end{align}
The  final master equations become 
\begin{equation}\label{Master}
 \Phi_\pm'' + \frac{(u^{1/2} f)'}{u^{1/2} f} \Phi_\pm' 
  + V_\pm \Phi_\pm = 0 , 
\end{equation}
where potentials $V_\pm$ are given by 
\begin{align}
 V_\pm
 &=
 \frac{1}{16u^2f^2(4b^2k^2-3f')^2}\times 
\notag\\&\quad
 \times\Bigg\{
 -4uf(4b^2k^2-3f')\Big(16b^2k^2(b^2k^2-C_\pm u+3au^2)
\notag\\&\hspace{12em}
 -4f'(2b^2k^2+3C_\pm u-9au^2) -3(f')^2
 \Big)
\notag\\&\qquad\quad
 +f^2
 \Big[
 16
 \Big(
 b^4k^4+12C_\pm b^2k^2u -108ab^2k^2u^2 +54C_\pm au^3-162a^2u^4
 \Big)
\notag\\&\qquad\qquad\qquad
 -24f'
 \Big(
 b^2k^2-6C_\pm u-18au^2
 \Big)
 +9(f')^2
 \Big]
 \Bigg\} 
 + \frac{b^2 \omega^2}{u f^2}. 
\end{align}

Let us solve these master equations. 
We first impose the incoming boundary condition at the horizon. 
Near the horizon $u=1$, 
the solutions of the master equation (\ref{Master}) behave as 
\begin{equation}
 \Phi_\pm \sim (1-u)^\nu , 
\end{equation}
where there are two independent solutions 
with $\nu = \pm i b \omega/(2-a)$. 
Each of them corresponds to the incoming and 
outgoing modes at the horizon. We impose the incoming boundary condition 
by taking only $\nu = -i b \omega/(2-a)$. 
Then, the solution can be factorized as 
\begin{equation}
 \Phi_\pm = (1-u)^{-ib \omega/(2-a)} F_\pm .
\end{equation}
Since we have factorized the most singular part near the horizon, 
$F_\pm$ should be regular at the horizon. 
In the hydrodynamic region, 
$\omega$ and $k$ are small and 
we can solve the master equations order by order by expanding 
the master fields with respect to $\omega$ and $k$.
It was solved in \cite{MSTTY}, 
and the solutions take the form of 
\begin{equation}
 \Phi_\pm = C_\pm H_\pm(u) (1-u)^{-ib\omega/(2-a)}
  \left(1 + b \omega F_{\pm 10}(u) + b^2 \omega^2 F_{\pm 20}(u) 
  + b^2 k^2 F_{\pm 02}(u) + \cdots \right) . 
\end{equation}
where $H_\pm$  given by 
\begin{align}
 H_+ &= u^{-3/4} , & 
 H_- &= \frac{u^{1/4}}{(1+a)-\frac{3}{2}au} , 
\end{align}
are factorized such that the leading terms in $F_{\pm}$ become $u$ independent constant 
. The determination of integration constants $C_\pm$ is important for computing the final Green functions and we will see it soon. 
Details of $F_{\pm ij}$ are shown in Appendix A. 

\subsection{Boundary Conditions at Slice $u=u_c$}
In order to determine the integration constant $C_\pm$ we need boundary conditions of the original fields
$h^\mu_\nu$ and $B_\mu$. 
Consider the boundary at slice $u=u_c$ and impose the boundary values for the original fields as 
\begin{align}
 h^\mu_\nu(u_c) &= {h^\mu_\nu}^{(0)} , & 
 B_\mu(u_c) &= {B_\mu}^{(0)} . 
\end{align}
Taking use of the Einstein equations and Maxwell equations at the slice, 
we obtain the following relations between 
the integration constants and boundary conditions: 
\begin{align}
 \ce_+ C_+ + \ce_- C_- &= \ce_0 , \\
 \cm_+ C_+ + \cm_- C_- &= \cm_0 , 
\end{align}
where $\ce_0$ and $\cm_0$ are 
linear combinations of ${h^\mu_\nu}^{(0)}$ and $(B_\mu)^{(0)}$.
Namely they take the form of 
\begin{align}
 \ce_0 &= \ce_{0I} \phi_I^{(0)} , & 
 \cm_0 &= \cm_{0I} \phi_I^{(0)} , 
\end{align}
where $\phi_I$ stands for the original fields, $h^\mu_\nu$ or $B_\mu$, 
and $\ce_{0I}, \cm_{0I}$ are given functions of 
$\omega$, $k$, $a$, $b$ and $u_c$. 
$\ce_\pm$ and $\cm_\pm$ are also known functions of 
$\omega$, $k$, $a$, $b$, $u_c$ and generally
they take the form of 
\begin{align}
 \ce_\pm &= \ce_{\pm 0} \widetilde F_\pm(u_c) + \ce_{\pm 1} \widetilde F'_\pm(u_c)  , \\
 \cm_\pm &= \cm_{\pm 0} \widetilde F_\pm(u_c) + \cm_{\pm 1} \widetilde F'_\pm(u_c)  ,
\end{align}
where $\ce_{\pm 0}, \ce_{\pm 1}, \cm_{\pm 0}, \cm_{\pm 1}$ are all known coefficients and $\widetilde F_\pm$ are defined in terms of the solutions 
of the master fields as 
\begin{equation}\label{tildeF}
 \widetilde F_\pm(u) = (1-u)^{-ib\omega/(2-a)}
  \left(1 + b \omega F_{\pm 10}(u) + b^2 \omega^2 F_{\pm 20}(u) 
  + b^2 k^2 F_{\pm 02}(u) + \cdots \right) . 
\end{equation}
Remember $\widetilde F_\pm(u)$ are given functions since we have solved the master equations. Once the above ingredients are given, $C_\pm$ can be finally expressed 
in terms of the boundary conditions as 
\begin{align}
 C_+ &= \frac{\ce_0 \cm_- - \cm_0 \ce_-}{\ce_+\cm_- - \cm_+\ce_-} , & 
 C_- &= \frac{\cm_0 \ce_+ - \ce_0\cm_+}{\ce_+\cm_- - \cm_+\ce_-} . 
\end{align}

\section{Pole of Green functions at the slice $u=u_c$}\label{pole}
Now we consider Green functions. 
By using the equations of motions, 
the on-shell action for fluctuations is reduced to surface terms as 
\begin{align}
 S
 &=
 \frac{l^3}{32\kappa^2b^4}\!\int\!\frac{d^4k}{(2\pi)^4}
 \Bigg\{
 \frac{1}{u}h^z_t{h^z_t}' +\frac{f}{u}h^x_x{h^x_x}'
 +\frac{f}{u}h^t_t{h^x_x}' +\frac{f}{2u}h^t_t{h^z_z}'
 \notag 
 \\
 &
 \qquad\qquad\qquad\qquad\quad
 +\frac{f}{u}h^x_x{h^t_t}' 
 +\frac{f}{u}h^x_x{h^z_z}'
 +\frac{f}{2u}h^z_z{h^t_t}'
 +\frac{f}{u}h^z_z{h^x_x}'
 \notag
 \\
 &
 \qquad\qquad\qquad\qquad\quad
 +\frac{3}{4u^2}\Big(f-\sqrt{f}\Big)(h^t_t)^2
 +\frac{1}{4u^2}\Big(3f-uf'-3\sqrt{f}\Big)(h^z_z)^2
 \notag
 \\
 &
 \qquad\qquad\qquad\qquad\quad
 -\frac{3}{u^2f}\Big(f-\sqrt{f}\Big)(h^z_t)^2
 -\frac{1}{2u^2}\Big(6f-uf'-6\sqrt{f}\Big)h^t_th^x_x
 \notag 
 \\
 &
 \qquad\qquad\qquad\qquad\quad
 -\frac{1}{4u^2}\Big(6f-uf'-6\sqrt{f}\Big)h^t_th^z_z
 -\frac{1}{u^2}\Big(3f-uf'-3\sqrt{f}\Big)h^x_xh^z_z
 \notag
 \\
 &
 \qquad\qquad\qquad\qquad\quad
 +3a 
 \Big(
 B_tB_t'-fB_zB_z'
 +\frac{1}{2}B_th^t_t
 +B_zh^z_t
 -B_th^x_x
 -\frac{1}{2}B_th^z_z
 \Big)
 \Bigg\}\Bigg|_{u=u_c}.  
\end{align}
It can be rewritten as 
\begin{equation}\label{osaction1}
 S = \int \frac{dk^4}{(2\pi)^2} 
  \Bigl(\phi_I(u_c) \ck_{IJ} \phi_J(u_c) 
   + \phi_I(u_c)\cl_{IJ}\phi_J'(u_c)\Bigr) . 
\end{equation}
By using the constraint equations and 
the definitions of the master fields, 
we obtain the relation between 
the original fields $\phi_I$ and its first order derivatives $\phi_I'$. 
Since we have 4 constraint equations and 2 master fields, 
we can diagonalize these equations such that
the 6 first derivatives $\phi_I'$ at the boundary 
can be expressed in terms of the boundary values $\phi_I^{(0)}$. 
They take the following form: 
\begin{align}\label{primefields}
 \phi_I'  
 &= 
 \ch_{IJ} \phi_J + \ch_{I+} C_+\widetilde F_+ + \ch_{I-} C_- \widetilde F_-\ .
\end{align}
Combined with (\ref{osaction1}) the on shell action is obtained as 
\begin{equation}
 S = \int \frac{dk^4}{(2\pi)^4} \phi_I^{(0)} \cg_{IJ} \phi_J^{(0)} , 
\end{equation}
where $\cg_{IJ}$ is given by 
\begin{align}\label{gfunction}
 \cg_{IJ} &= 
 \ck_{IJ} + \cl_{IK} 
 \biggl(\ch_{KJ} 
 + \ch_{K+}\tf_+(u_c) 
 \frac{\ce_{0J} \cm_- - \cm_{0J} \ce_-}{\ce_+\cm_- - \cm_+\ce_-} 
\notag\\&\hspace{10em}
 + \ch_{K-}\tf_-(u_c)
 \frac{\cm_{0J} \ce_+ - \ce_{0J}\cm_+}{\ce_+\cm_- - \cm_+\ce_-}
 \biggr) . 
\end{align}
Finally the retarded Green functions at the slice $G^\mathrm{R}$ are given by~\cite{Son.Realgreenfunction} 
\begin{equation}
 G^\mathrm{R}_{IJ} = 2 \cg_{IJ}\ .
\end{equation}
We find the pole of the Green functions is located at
\begin{equation}
 \ce_+ \cm_- - \ce_- \cm_+ = 0 . 
\end{equation}
For convenience we rewrite (\ref{tildeF}) as 
\begin{align}
 \tf_\pm(u) = 
 1 + b \omega \tf_{\pm10}(u) + b^2 \omega^2 \tf_{\pm20}(u) 
 + b^2 k^2 \tf_{\pm02}(u) + \cdots. 
\end{align}
The pole has the following structure: 
\begin{align}\label{pole1}
 0 &= 
  P_{40} \omega^4 + P_{30} \omega^3 
 + P_{22} \omega^2 k^2 + + P_{12} \omega k^2
 + P_{04} k^4 + \cdots\ ,
\end{align} where the coefficients are listed as follows
\bea
P_{40} &=&   9 b^2 f'(u_c) 
 \left[\tf_{+10}\tf'_{-10} + \tf'_{-20}(u_c)\right] \ ,\\
P_{30} &=&  9 b f'(u_c) \tf'_{-10}(u_c) \ ,\\
P_{22} &=&  \biggl\{
 - 3 b^2 \left[f(u_c) (f'(u_c)+ 6 a u_c^2) 
 - u_c f^{\prime\,2}(u_c)\right]
 \left[\tf_{+10}\tf'_{-10} + \tf'_{-20}(u_c)\right]\\
 & &+ 9 b^2 f'(u_c) \tf'_{-02}(u)
 - 12 b^2 f^2(u_c) \tf'_{+10}(u_c)\tf'_{-10}(u_c) 
 +\frac{54 a^2 b^2 u_c^2}{(1+a) f'(u_c)}
 \biggr\}\ ,\\
P_{12} &=&   - 3 b \left[f(u_c) (f'(u_c)+ 6 a u_c^2) 
 - u_c f^{\prime\,2}(u_c)\right] \tf'_{-10}(u_c)\ ,\\
P_{04} &=&  \Bigl\{- 3 b^2 \left[f(u_c) (f'(u_c)+ 6 a u_c^2) 
 - u_c f^{\prime\,2}(u_c)\right] \tf'_{-02}(u_c)
 + \frac{6 a b^2 u_c^3}{(1+a) f^{\prime\,2}(u_c)}\times \\ 
 & &\Bigl[ 6 a (1 + a) + (8 + 24 a-3 a^2 + 8 a^3) u_c 
 - 18 a (1 + a)^2 u_c^2 + 15 a^2 (1 + a) u_c^3\Bigr]
 \Bigr\}\ .
\eea Apparently $P_{ij}$ coefficients here can only be fixed up to an overall constant and we leave the explicit form of $\ce_+ \cm_- - \ce_- \cm_+$ to the Appendix B.
Due to the structure (\ref{pole1}), the pole has 4 roots which are referred to as 
\begin{align}
 \omega &= 1/\tau + \cdots , & 
 \omega &= \pm c_s k +\cdots , & 
 \omega &= - i D_A k^2 + \cdots, 
\end{align} 
and can be factorized to be product of these roots:  
\begin{align}
 0 &= (\omega - c_s k + \cdots)(\omega + c_s k + \cdots)
 (\omega + i D_A k^2 + \cdots)(\omega - 1/\tau + \cdots) \\
 &= -\frac{1}{\tau}\omega^3 + \omega^4 
 + \frac{c_s^2}{\tau}\omega k^2 
 - \left(c_s^2 - i \frac{D_A}{\tau}\right) \omega^2 k^2 
 -i \frac{c_s^2 D_A}{\tau} k^4 + \cdots\ .\label{pole2}
\end{align} Here ``$\cdots$'' denote the higher derivative corrections.
By matching (\ref{pole1}) and (\ref{pole2}), 
we obtain the bare sound speed
\begin{align}
 c_s^2 
 &= - \frac{P_{12}}{P_{30}} 
= 
 \frac{f(u_c) [f'(u_c)+ 6 a u_c^2] 
 - u_c f^{\prime\,2}(u_c)}{3f'(u_c)}\ ,
 \end{align} and the bare charge diffusion constant
 \begin{align}
 D_A 
 &= - i \frac{P_{04}}{P_{12}}
 = \mathrm{Im} \Biggl[ b \frac{\tf'_{-02}(u_c)}{\tf'_{-10}(u_c)} 
\notag\\&\quad
 - b \frac{2 a u_c^3[ 6 a (1 + a) + (8 + 24 a - 3 a^2 + 8 a^3) u_c 
 - 18 a (1 + a)^2 u_c^2 + 15 a^2 (1 + a) u_c^3]}
 {(1+a) f^{\prime\,2}(u_c)\left[f(u_c) (f'(u_c)+ 6 a u_c^2) 
 - u_c f^{\prime\,2}(u_c)\right] \tf'_{-10}(u_c)}\Biggr]\ .
\end{align} In the orthonormal frame,
the proper frequency $\omega_c$ and proper momentum $k_c$ 
are defined by  
\begin{align}
 \omega_c &= \frac{\omega}{\sqrt{-g_{tt}}}\ , & 
 k_c &= \frac{k}{\sqrt{g_{zz}}}\ .
\end{align}
The normalized speed of sound and the diffusion constant are given by
\begin{align}
 \bar c_{s}^2 
 &= - \frac{g_{zz}}{g_{tt}} c_s^2 
= \frac{f(u_c) [f'(u_c)+ 6 a u_c^2] 
 - u_c f^{\prime\,2}(u_c)}{3 f(u_c) f'(u_c)} , \\
 \bar D_{A}  &= - \frac{g_{zz}}{g_{tt}} T_H D_A 
= \mathrm{Im}\Biggl[\frac{2-a}{4\pi}\frac{\tf'_{-02}(u_c)}{f(u_c) \tf'_{-10}(u_c)} 
\notag\\&\quad
 - \frac{a (2-a) u_c^3 [ 6 a (1 + a) + (8 + 24 a - 3 a^2 + 8 a^3) u_c 
 - 18 a (1 + a)^2 u_c^2 + 15 a^2 (1 + a) u_c^3]}
 {2\pi (1+a) f(u_c) f^{\prime\,2}(u_c)\left[f(u_c) (f'(u_c)+ 6 a u_c^2) 
 - u_c f^{\prime\,2}(u_c)\right] \tf'_{-10}(u_c)}\Biggr]\ .
\end{align}

\section{RG  Flows of sound speed and diffusion constants}
\begin{figure}[hbtp]
\includegraphics*[bb=0 0 250 160,width=0.45\columnwidth]{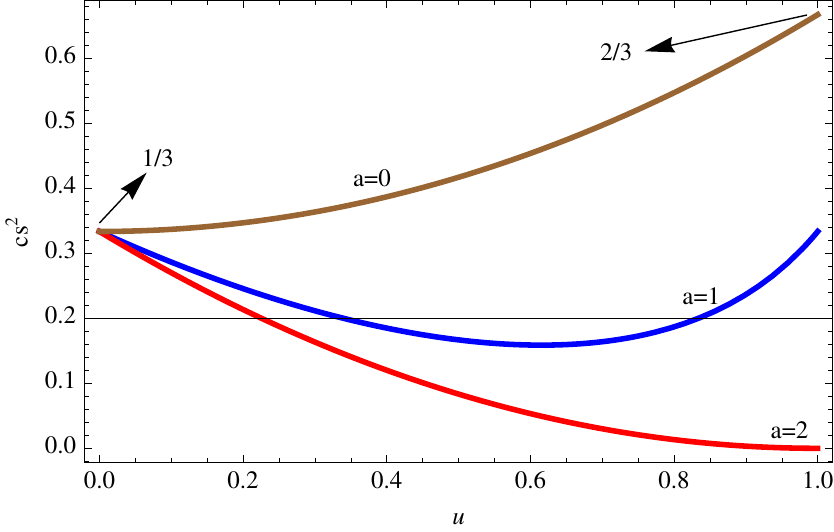}~~~~~~~~~
\includegraphics*[bb=0 0 250 160,width=0.45\columnwidth]{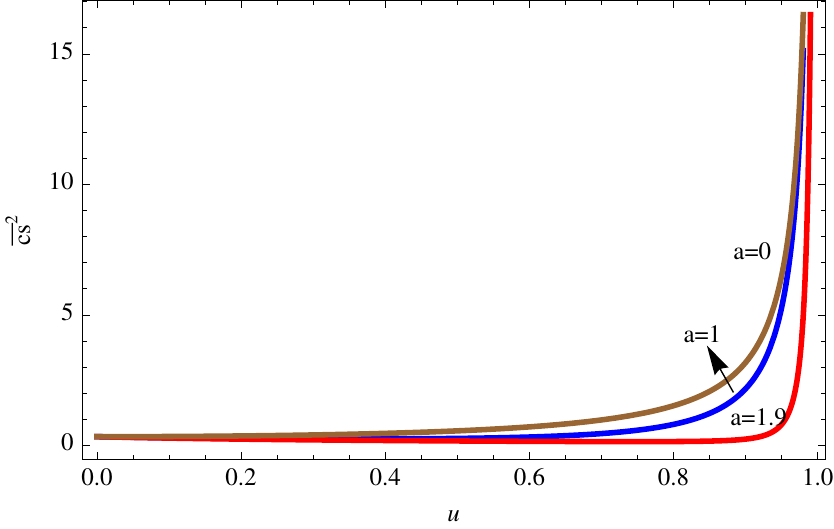}
\caption{ Radial flow of speed of sound $c_s^2$. Left: The bare speed of sound. Right: Normalized speed of sound in orthonormal frame.
  \label{fig1}}
\end{figure}

\begin{figure}[hbtp]
\includegraphics*[bb=0 0 250 160,width=0.5\columnwidth]{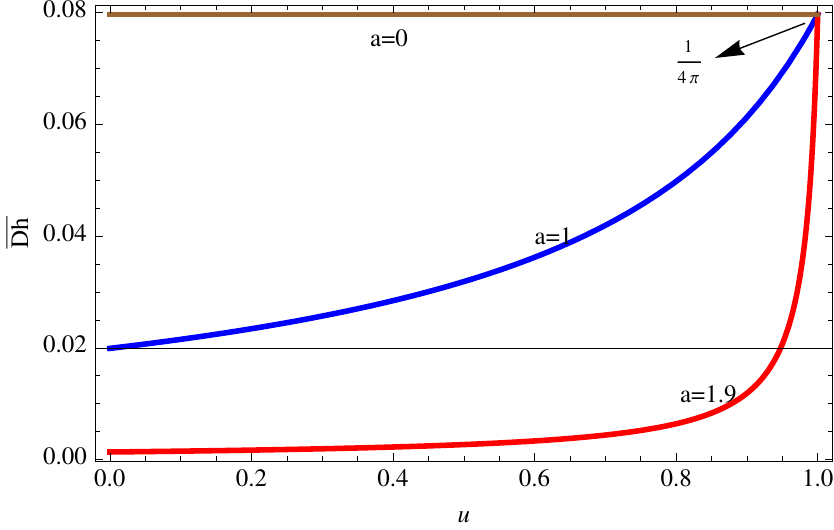}~~~~~~~~~~~~~
\includegraphics*[bb=0 0 280 170,width=0.45\columnwidth]{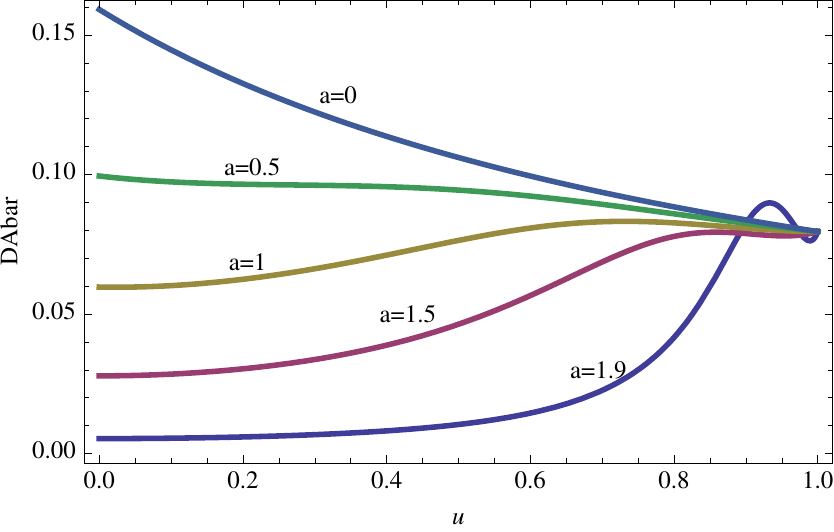}
\caption{ Radial flow of diffusion constants. Left: The momentum diffusion constant coming from shear part hydrodynamical pole~\cite{MSZ}. Right: Charge diffusion constant coming from sound part hydrodynamical pole.
  \label{fig2}}
\end{figure}

We will summarize radial flows of various quantities in this section. 
It is shown in Left part of Fig \ref{fig1} that the bare speed of sound flows quite differently for different charges. For zero charge case, $c_s^2(u)$ runs from ${1\over 3}$ to ${2\over 3}$ if we move from UV boundary to IR horizon. This is consistent with the chargeless result in~\cite{Andy.RG,Marolf:2012dr} at the horizon and also at the UV boundary~\cite{Kovtun:2005ev}. As suggested in~\cite{Andy.RG}, the physical speed of sound may be defined as the one in the orthonormal frame, which is shown in right part of Fig \ref{fig1}.
The normalized sound speed always diverges at the horizon for any charge. One interesting observation is that both for bare one or normalized one, there is a universal UV boundary value of sound speed ${1\over \sqrt{3}}$ for different charges. Let us turn to the diffusion. Remember the momentum diffusion $D_h$ is encoded in hydrodynamical pole in shear part of metric fluctuations~\cite{MSZ} and charge diffusion $D_A$ is contained in sound part. As shown in Fig \ref{fig2}, both for normalized $\bar D_h$ and $\bar D_{A}$, the universal charge independent value ${1\over 4\pi}$ is located at the horizon. 

\vspace{.5cm}

\vspace{.5cm}

\section{Transports and Einstein relation at the slice $r_c$}\label{transport}
Retarded Green functions can be computed explicitly following (\ref{gfunction}).
Generally the Green functions have complicated expressions. Rather than showing all the explicit expressions we focus on  
$G_{z\,z}$, $G_{ii\,jj}$ and $G_{t\,t}$ at certain limits.
By using the Kubo formula, we obtain 
\begin{align}
 \sigma &= 
 \lim_{\omega\to 0} \frac{1}{\omega}\mathrm{Im} G_{z\,z}(\omega,k=0) 
 = \frac{l^3}{16 \kappa^2 b^4 \mu^2} 
 \frac{3 a b}{2 u_c f(u_c) \mathrm{Im}\tf_{-10}'(u_c)} , \notag\\
 &= \frac{l}{2 e^2 b u_c f(u_c) \mathrm{Im}\tf_{-10}'(u_c)}\ .
 \end{align}
 This is electrical conductivity, which is the same as our previous result~\cite{MSZ}. We also find that the bulk viscosity vanishes at arbitrary slice $u=u_c$
 \begin{align}
 \zeta &= 
 \lim_{\omega\to 0} \frac{1}{\omega}
 \sum_{i,j=x,y,z}\mathrm{Im} G_{ii\,jj}(\omega,k=0) 
 = 0\ .
\end{align}
In $k\to 0$ limit, the density-density correlation function $G_{t\,t}$ behaves as 
\begin{equation}
 \lim_{k\to 0} \frac{\mathrm{Im} G_{t\,t}}{k^2} = \frac{l^{3}}{16 b^4 \kappa^2 \mu^2}
 \frac{3 a b}{2 \omega u_c f(u_c) \mathrm{Im}\tf_{-10}'(u_c)} . 
\end{equation}
Together with the pole structure which we have discussed above, 
it implies that 
\begin{equation}
 \mathrm{Im} G_{t\,t}(\omega,k) \sim 
 \frac{l^3}{16 b^4 \kappa^2 \mu^2}
 \frac{3 a b}{2 u_c f(u_c) \mathrm{Im}\tf_{-10}'(u_c)} 
 \frac{k^2 \omega^3}{(\omega^2 + D_A^2 k^4)(\omega^2 - c_s^2 k^2)} + \cdots
\end{equation}
and then we obtain 
\begin{equation}
 \mathrm{Im} G_{t\,t} (\omega,k=0) = 
 \lim_{k\to 0} \mathrm{Im} G_{t\,t} (\omega,k) = 
 \frac{l}{2 e^2 b}
 \frac{-\omega}{u_c f(u_c) \mathrm{Im}\tf_{-10}'(u_c)} 
 \frac{2\pi}{D_A} \delta(\omega)\ .
\end{equation}
By using the definition of the charge susceptibility 
\begin{equation}
 \Xi = - \frac{1}{T} \int \frac{d \omega}{2\pi} \frac{\mathrm{Im}G_{t\,t}(\omega,k=0)}{e^{\omega/T}-1} , 
\end{equation}
we obtain 
\begin{equation}
 \Xi = \frac{1}{D_A} 
 \frac{l}{2 e^2 b u_c f(u_c) \mathrm{Im}\tf_{-10}'(u_c)} , 
\end{equation}
which satisfies the Einstein relation 
\begin{equation}
 \Xi = \frac{\sigma}{D_A} . 
\end{equation}
 As a summary, for the transport coefficients, the shear viscosity keeps a constant from the horizon to the UV boundary and DC electric conductivity has a nontrivial radial flow as explicitly shown in~\cite{MSZ}, . Here we checked that the bulk viscosity vanishes at arbitrary slice and the Einstein relation holds at the slice.

\section*{Acknowledgements}
This work was supported by the National Research Foundation of Korea(NRF) grant funded by the
Korea government(MEST) through the Center for Quantum Spacetime(CQUeST) of Sogang University with grant number
2005-0049409. SJS was also supported by Mid-career Researcher Program through NRF grant (No. 2010-0008456 ). 
YM is supported by JSPS Research Fellowship for Young Scientists and 
in part by Grant-in-Aid for JSPS Fellows (No.23-2195). 

\appendix 

\section{Concrete expressions}

The solutions of $F_\pm$ are the following: 
\begin{subequations}
\begin{align}
F_{+10}(u)
&=
\frac{i}{2 (2-a)}
\left\{\log \left(1+u-au^2\right)
-\frac{6 K_1(u)}{\sqrt{1+4a}}
\right\},
\\
F_{+02}(u)
&=
\frac{2}{3}  
\left\{\frac{K_1(u)}{\sqrt{1+4a}}-\frac{1}{(1+a)u}
\right\}, 
\\
F_{+20}(u)
&=\int^u d u\frac{1}{(1-u)(1+u-au^2)} 
\notag\\&\qquad\qquad
\times
\Biggl\{1-u+\frac{(1-u) (1+au) \log \left(1+u-a u^2\right)}{2 (2-a)^2}
-\frac{3 (1-u) (1+au) K_2(0)}{2 (2-a)^2 \sqrt{1+4a}}
\notag\\
&\qquad\qquad\qquad
-\frac{(1+a) K_2(1) u}{\sqrt{1+4a}}
+\frac{\Big(3+(5+3a-6a^2+2a^3)u-3au^2\Big)
   K_2(u)}{2 (2-a)^2 \sqrt{1+4a}}\Biggr\}, \\
F_{-10}(u)
&=
\frac{i}{2 (2-a)^2}
\Bigl\{
8(1+a)^2 \log (u)-(2+a) (1+4a) 
\log\Big(1+u-au^2\Big)
\notag\\&\qquad\qquad\qquad\qquad\qquad\qquad
-2 \sqrt{1+4a} (2+5a) K_1(u)
\Bigr\}, 
\\ 
F_{-02}(u)
&=
\Biggl\{-\frac{3a^2u}{2(1+a)^2\Big(\displaystyle 1+a-\frac{3}{2}au\Big)}
-\frac{2 (1+a) (2+a) \log (u)}{(2-a)^2} 
\notag\\
&\qquad\qquad\qquad
+\frac{(1+a)(2+a) 
\log\Big(1+u-au^2\Big)}{(2-a)^2}+\frac{2 (2+5a+6a^2) K_1(u)}{(2-a)^2 
\sqrt{1+4a}}
\Biggr\}, 
\\
F_{-20}(u)
&=
\int^u d u
\frac{1}{2 (2-a)^4 (1+4a)^{3/2} (1-u) u 
\left(1+u-a u^2\right)}
\notag\\
&\qquad\times
\Biggl\{8 (2-a)(1+a)^2(1+4a)^{3/2} u (1+u-au^2)\log (u)
\notag\\
&\qquad\qquad
-(2-a) (1+4a)^{3/2} 
\Bigl(
4(1+a)^2
+(2-3a-8a^2)u
+(2+9a+13a^2)u^2
\notag\\
&\qquad\qquad\qquad\qquad\qquad\qquad\qquad\qquad\qquad
-a (2+a) (1+4a)u^3
\Bigr) 
\log \Big(1+u-a u^2\Big)
\notag\\
&\qquad\qquad
+(1+4a)^2 (2+5a)K_2(0) 
(1-u) 
\Bigl(4(1+a)^2+(2-3a-8a^2)u
-(2-a)au^2\Bigr)
\notag\\
&\qquad\qquad
-2(1+a)(2-2a+41a^2)K_2(1)\Big(2(1+a)-3au\Big)^2
\notag\\
&\qquad\qquad
-(2-a)
\Bigl(a(1+4a)^2(2+5a)u^3
-(2-a)(1+11a+46a^2+18a^3)u^2
\notag\\
&\qquad\qquad\qquad\qquad\qquad
-\left(2+9a+180a^2+224a^3+24a^4\right)u
\notag
\\&\qquad\qquad\qquad\qquad\qquad\qquad
-4(1+a)^2 (1-10a-2a^2)
\Bigr)K_2(u)
\Biggr\}.
\end{align}
\end{subequations}
where
\begin{align}
K_1(u)
&=\frac{1}{2}\log(1+u-au^2)-\log\left(1-\frac{2au}{1+\sqrt{1+4a}}\right),
\\
K_2(u)
&=\log\left(\frac{1+\sqrt{1+4a}-2au}{-1+\sqrt{1+4a}+2au}\right). 
\end{align}
Then, $\tf_\pm$ are related to these solutions as 
\begin{align}
 \tf_{\pm 10}(u) 
 &=
 F_{\pm 10} - \frac{i}{2-a}\log(1-u) 
\\
 \tf_{\pm 20}(u)
 &= 
 F_{\pm 20}(u) - \frac{i}{2-a}\log(1-u) F_{\pm 10}(u) 
 - \frac{1}{2(2-a)^2}\log^2(1-u)
\\
 \tf_{\pm 02}(u) 
 &= 
 F_{\pm 02}(u) 
\end{align}

$\ce_{0I}$ and $\cm_{0I} $are defined as 
coefficients of the original fields in 
\begin{align}
 \ce_0 &= \ce_{0I}\phi_I^{(0)} , & 
 \cm_0 &= \cm_{0I}\phi_I^{(0)} . 
\end{align}
They can be read off from the expressions of 
$\ce_0$ and $\cm_0$: 
\begin{align}
 \ce_0 
 &= 
 \frac{1}{24 k^2 u_c f^3(u_c)}
 \biggl\{
 -k u_c f(u_c) (4 b^2 k^2 - 3 f'(u_c)) 
 \Bigl[
 -4 b^2 k^3 (-\httc + \hxxc) 
\notag\\&\qquad
 - 18 a u_c (k \btc - k u_c \hxxc + \omega \bzc) 
 + 3 k f'(u_c)(-\httc + \hxxc) 
 \Bigr] 
\notag\\&\qquad
 + u_c (4 b^2 k^2 - 3 f'(u_c))^2 
 \left[
 2 \omega k \htzc - (\omega^2+ k^2 u_c f'(u_c)) \hxxc + \omega^2 \hzzc 
 \right]
 \biggr\} , \\
 \cm_0 
 &= 
 \frac{1}{8k^2 u_c f^2(u_c)}
 \biggl\{
 k f(u_c) 
 \Bigl[
 -18 a u_c^2 (k \btc - k u_c \hxxc + \omega \bzc) 
\notag\\&\qquad
 -4 b^2 k^2 ( 2 k \btc - k u_c\httc - k u_c \hxxc + 2 \omega \bzc) 
 + 3 k u_c f'(u_c) (\hxxc - \httc)
 \Bigr]
\notag\\&\qquad
 + u (4 b^2 k^2 - 3 f'(u_c)) 
 \left[
  - 2 \omega k \htzc + (\omega^2+ k^2 u_c f'(u_c)) \hxxc - \omega^2 \hzzc 
 \right]
 \biggr\} . 
\end{align}
$\ce_\pm$ and $\cm_\pm$ 
contain $\tf_\pm(u_c)$ and their derivatives. 
They are referred to as 
$\ce_{\pm 0}$, $\ce_{\pm 1}$, 
$\cm_{\pm 0}$ and $\cm_{\pm 1}$, 
respecctively, and obtained as 
\begin{align}
 \ce_{+0}
 &= 
 \frac{1}{12 (1+a)\alpha u_c f^3(u_c) k^2}
 \biggl\{
 -3 k^2 u_c f(u_c) (4b^2 k^2 - 3 f'(u_c)) 
 (4 b^2 k^2 +2 u_c \alpha_- - f'(u_c)) 
\notag\\&\hspace{12em}
 + u_c (4b^2 k^2 - 3 f'(u_c))^2 ( 3 \omega^2 + u_c f'(u_c) k^2) 
\notag\\&\hspace{12em}
 - 12 k^2 f^2(u_c) (4 b^2 k^2 + 3(1+a)(1-\alpha)u_c)
 \biggr\}
\\
 \ce_{+1}
 &= 
 \frac{4 b^2 k^2 - 3 f'(u_c)}{(1+a)\alpha f(u_c)}
\\
 \ce_{-0}
 &= 
 \frac{u_c}{6(1+a)\alpha k^2 f^3(u_c) f^{\prime\,2}(u_c)}
 \times
\notag\\&\quad\times
 \biggl\{ 
 2 k^2 f^{\prime\,2}(u_c) 
 \Bigl[
 \left(3 u_c f(u_c) (3\alpha_+ u_c+8 b^2 k^2)-18 f^2(u_c)\right)
 +4 b^2 u_c (2b^2 k^4 u_c - 9 \omega^2)
 \Bigr]
\notag\\&\qquad
 -12 k^2 f'(u_c) 
 \Bigl[2 b^2 k^2
 u_c f(u_c) (\alpha_+ u_c+2 b^2 k^2)
 +f^2(u_c) \left(3 \alpha_+ u_c-4 b^2 k^2\right)
 -4 b^4 k^2 \omega ^2 u_c
 \Bigr]
\notag\\&\qquad
 -48 b^2 k^4 u_c f^2(u_c) f''(u_c) 
 -3 u_c f^{\prime\,3}(u_c)
 \left(8 b^2 k^4 u_c+3 k^2 f(u_c)-9 \omega^2\right) 
 + 9 k^2 u_c^2 f^{\prime\,4}(u_c)
 \biggr\}
\\
 \ce_{-1} 
 &=
 \frac{2 u_c^2 (4b^2 k^2 - 3 f'(u_c))}
 {(1+a)\alpha f(u_c) f'(u_c)}
\\
 \cm_{+0}
 &= 
 - \frac{1}{4 a (1+a) \alpha u_c^2 k^2 f^2(u_c)} 
 \biggl\{
 4 (1+a) (1-\alpha) f^2(u_c) k^2 
\notag\\&\hspace{12em}
 + a u_c^2 f(u_c) k^2 (-4 b^2 k^2 - 6 u_c \alpha_- + 3 f'(u_c)) 
\notag\\&\hspace{12em}
 + a u_c^2 ( 4 b^2 k^2 -3 f'(u_c))(3\omega^2 + u_c f'(u_c)k^2) 
 \biggr\}
\\
 \cm_{+1} 
 &= 
 \frac{\alpha_-}{a (1+a) u_c \alpha} 
\\
 \cm_{- 0}
 &= 
 \frac{1}{2a(1+a)\alpha k^2 f^{\prime\,2}(u_c) f^2(u_c)} 
 \biggl\{
 - 4 k^2 \alpha_+ u_c f''(u_c) f^2(u_c)
\notag\\&\qquad
 + 2 k^2 f'(u_c) 
 \Bigl[
 -6 a u_c^2 b^2 \omega^2 
 + a u_c^2 ( 2 b^2 k^2 + 3 u_c \alpha_+) f(u_c) 
 + 2 (\alpha_+ - 3 a u_c) f^2(u_c) 
 \Bigr]
\notag\\&\qquad
 - a u_c^2 f^{\prime\,2}(u_c) 
 \left[4 u_c b^2 k^4 - 9 \omega^2 + 3 k^2 (f(u_c)- u_c f'(u_c))\right]
 \biggr\}
\\
 \cm_{-1}
 &= 
 \frac{2 \alpha_+ u_c}{a (1+a) \alpha f'(u_c)} . 
\end{align}
The first derivatives of original fields are expressed as 
\begin{align}
 \phi_I'  
 &= 
 \ch_{IJ} \phi_J + \ch_{I+} C_+\tf_+ + \ch_{I-} C_- \tf_- ,  
\end{align}
whose explicit forms are as follows: 
\begin{align}
 {h^t_t}' 
 &= 
 - \frac{2 b^2 k^2}{3 f} h^t_t 
 + \frac{2 b^2 \omega^2}{3 f^2} h^z_z 
 + \frac{4 b^2 k\omega}{3 f^2} h^z_t 
\notag\\&\quad
 + \frac{f \left(-18 a u^2+4 b^2 k^2-9 f'\right)
 -4 b^2 k^2 u f'+8 b^2 \omega
   ^2+3 u {f'}^2}{6 f^2} h^x_x 
\notag\\&\quad
 -\frac{3 f \left(4 b^2 k^2-3 f'+2 u \alpha _-\right)+u f' \left(3 f'-4 b^2
 k^2\right)}{3 (a+1) \alpha  f^2}
 C_+ \tf_+ 
\notag\\&\quad
 -\frac{2 u^2 \left(3 f \left(4 b^2 k^2-3 f'+2 u \alpha _+\right)+u f' \left(3 f'-4
 b^2 k^2\right)\right)}{3 (a+1) \alpha  f^2 f'}
 C_- \tf_- 
\\
 {h^x_x}' 
 &= 
 \frac{4 b^2 k^2-3 f'}{12 f} h^t_t 
 + \frac{\omega ^2 \left(3 f'-4 b^2 k^2\right)}{12 f^2 k^2} h^z_z 
 + \frac{3 \omega  f'-4 b^2 k^2 \omega }{6 f^2 k} h^z_t 
\notag\\&\quad
 + \frac{f k^2 \left(18 a u^2-4 b^2 k^2+9 f'\right)+\left(4 b^2 k^2-3 f'\right)
   \left(k^2 u f'+\omega ^2\right)}{12 f^2 k^2} 
 h^x_x 
 -\frac{3 a u}{2 f} B_t 
 -\frac{3 a u \omega }{2 f k} B_z 
\notag\\&\quad
 -\frac{\left(4 b^2 k^2-3 f'\right) \left(k^2 u f'+3 \omega ^2\right)-3 f k^2 \left(4
   b^2 k^2-3 f'+2 \alpha _- u\right)}{6 (a+1) \alpha  f^2 k^2} C_+ \tf_+ 
\notag\\&\quad
 -\frac{u^2 \left(\left(4 b^2 k^2-3 f'\right) \left(k^2 u f'+3 \omega ^2\right)-3 f
   k^2 \left(4 b^2 k^2-3 f'+2 \alpha _+ u\right)\right)}{3 (a+1) \alpha
 f^2 k^2 f'} C_- \tf_- 
\\
 {h^z_z}' 
 &= 
 \frac{3 f'-4 b^2 k^2}{6 f} h^t_t 
 + \frac{\omega ^2 \left(4 b^2 k^2-3 f'\right)}{6 f^2 k^2} h^z_z 
 + \frac{\omega  \left(4 b^2 k^2-3 f'\right)}{3 f^2 k} h^z_t 
\notag\\&\quad
 -\frac{2 f \left(9 a k^2 u^2+4 b^2 k^4\right)+\left(4 b^2 k^2-3 f'\right) \left(k^2 u
   f'+\omega ^2\right)}{6 f^2 k^2} h^x_x 
 + \frac{3 a u}{f} B_t 
 + \frac{3 a u \omega }{f k} B_z 
\notag\\&\quad
 + \frac{\left(4 b^2 k^2-3 f'\right) \left(k^2 u f'+3 \omega ^2\right)-6 \alpha _- f k^2
   u}{3 (a+1) \alpha  f^2 k^2}C_+ \tf_+ 
\notag\\&\quad
 + \frac{2 u^2 \left(\left(4 b^2 k^2-3 f'\right) \left(k^2 u f'+3 \omega ^2\right)-6
 \alpha _+ f k^2 u\right)}{3 (a+1) \alpha  f^2 k^2 f'}
 C_- \tf_- 
\\
 {h^z_t}' 
 &= 
 \frac{\omega  \left(4 b^2 k^2-f'\right)}{2 f k} h^x_x 
 + \frac{\omega  f'}{2 f k} h^z_z 
 + \frac{f'}{f} h^z_t 
\notag\\&\quad
 -\frac{4 b^2 k^2 \omega -3 \omega  f'}{a \alpha  f k+\alpha  f k} C_+ \tf_+ 
 -\frac{2 u^2 \omega  \left(4 b^2 k^2-3 f'\right)}{(a+1) \alpha  f k f'}
  C_- \tf_- 
\\
 B_t' 
 &= 
 - \frac{1}{2} h^t_t 
 + \frac{3}{2} h^x_x
 + \frac{\alpha _-}{a (a+1) \alpha  u} C_+ \tf_+ 
 + \frac{2 u^2\alpha _+}{a (a+1) \alpha f'} C_- \tf_- 
\\
 B_z' 
 &= 
 -\frac{\omega }{2 f k} h^x_x
 + \frac{\omega }{2 f k} h^z_z 
 + \frac{1}{f} h^z_t 
 -\frac{\alpha _- \omega }{a (a+1) \alpha  f k u} C_+ \tf_+ 
 -\frac{2 \alpha _+ u \omega }{a (a+1) \alpha k f f'} C_- \tf_- 
\end{align}

\section{Pole Structure}
The pole is expanded as 
\begin{align}
 \frac{u_c}{a (1+a) f^3(u_c)}\Biggl\{& 
 9 b f'(u_c) \tf'_{-10}(u_c) \omega^3 
 + 9 b^2 f'(u_c) 
 \left[\tf_{+10}\tf'_{-10} + \tf'_{-20}(u_c)\right] \omega^4 
\notag\\&\quad
 - 3 b \left[f(u_c) (f'(u_c)+ 6 a u_c^2) 
 - u_c f^{\prime\,2}(u_c)\right] \tf'_{-10}(u_c)\omega k^2 
\notag\\&\quad
 + \biggl\{
 - 3 b^2 \left[f(u_c) (f'(u_c)+ 6 a u_c^2) 
 - u_c f^{\prime\,2}(u_c)\right]
 \left[\tf_{+10}\tf'_{-10} + \tf'_{-20}(u_c)\right]
\notag\\&\qquad\qquad
 + 9 b^2 f'(u_c) \tf'_{-02}(u)
 - 12 b^2 f^2(u_c) \tf'_{+10}(u_c)\tf'_{-10}(u_c) 
\notag\\&\qquad\qquad\qquad\qquad
 +\frac{54 a^2 b^2 u_c^2}{(1+a) f'(u_c)}
 \biggr\}\omega^2 k^2 
\notag\\&\quad
 + \Bigl\{- 3 b^2 \left[f(u_c) (f'(u_c)+ 6 a u_c^2) 
 - u_c f^{\prime\,2}(u_c)\right] \tf'_{-02}(u_c) 
\notag\\&\qquad\qquad
 + \frac{6 a b^2 u_c^3}
 {(1+a) f^{\prime\,2}(u_c)}
 \Bigl[
 6 a (1 + a) + (8 + 24 a - 3 a^2 + 8 a^3) u_c 
\notag\\&\qquad\qquad\qquad\qquad
 - 18 (a (1 + a)^2) u_c^2 + 15 a^2 (1 + a) u_c^3
 \Bigr]
 \Bigr\} k^4 + \cdots
 \Biggr\} 
\notag\\
 & = 0 . 
\end{align}

\end{document}